\documentclass[prd, preprint,nofootinbib,superscriptaddress,12pt]{revtex4-1}
\pdfoutput=1

\newcount\showkey \showkey=0  
\newcount\draft \draft=1 

\usepackage[dvipsnames]{xcolor}
\usepackage{tabularx}

\ifnum\showkey=1
  \usepackage{seqsplit}
  \usepackage[color]{showkeys}
    \definecolor{refkey}{rgb}{0.9, 0.43, 0.63}
    \definecolor{labelkey}{rgb}{0.59, 0.43, 0.63}
  \usepackage{xstring}
  
\fi

\usepackage[pdfusetitle,pdfborder={0 0 0}]{hyperref}
   \hypersetup{colorlinks,
               linkcolor={red!70!black},
               citecolor={green!40!black},
               urlcolor={blue!80!black}}
\usepackage{microtype}   
\usepackage{amsmath}     
  \usepackage{amsthm}    
  \usepackage{amssymb}   
\usepackage{siunitx}     
\usepackage{booktabs}    
\usepackage[capitalize]{cleveref}    
  
  \crefname{section}{Sec.}{Secs.}
  \crefname{appendix}{App.}{Apps.}

\usepackage[shortlabels,inline]{enumitem}  
\usepackage{mathtools} 
\usepackage{subdepth}    

\usepackage{graphicx,multirow,subfigure}
\usepackage{float} 
\usepackage[T1]{fontenc}
\usepackage{physics}
\usepackage{bbold}
\usepackage{rotating} 
\setlist[enumerate,2]{leftmargin=0.45em}
\usepackage{comment}

\DeclareFontFamily{U}{mathx}{\hyphenchar\font45}
\DeclareFontShape{U}{mathx}{m}{n}{<-> mathx10}{}
\DeclareSymbolFont{mathx}{U}{mathx}{m}{n}
\DeclareMathAccent{\widebar}{0}{mathx}{"73}

\newcommand{\beq}{\begin{equation}}
\newcommand{\eeq}{\end{equation}}

\ifnum\draft=1
  \newcommand{\ygn}[1]{{\bf \color{orange} #1}}
 \else
  \newcommand{\ygn}[1]{}
  \newcommand{\JBn}[1]{}
  \newcommand{\MFn}[1]{}
  \newcommand{\sts}[1]{}
  \newcommand{\JZn}[1]{}
\fi

\definecolor{nicered}{rgb}{0.7,0.1,0.1}
\definecolor{nicegreen}{rgb}{0.1,0.5,0.1}
\hypersetup{colorlinks,citecolor= blue,linkcolor= nicered}

\newenvironment{Eqnarray}{\arraycolsep 0.14em\begin{eqnarray}}{\end{eqnarray}}
\def\beqa{\begin{Eqnarray}}
\def\eeqa{\end{Eqnarray}}
\newcommand{\no}{\nonumber}

\newcommand{\bea}{\begin{eqnarray}}
\newcommand{\eea}{\end{eqnarray}}

\def\lsim{\mathrel{\rlap{\lower4pt\hbox{\hskip1pt$\sim$}}
     \raise1pt\hbox{$<$}}}         
\def\gsim{\mathrel{\rlap{\lower4pt\hbox{\hskip1pt$\sim$}}
     \raise1pt\hbox{$>$}}}         

\begin{document}

\title{The branching fraction of $B_{s}^0\to K^0\overline{K}{}^0$: Three puzzles}

\author{Yasmine~Amhis}
\email{yasmine.sara.amhis@cern.ch}
\affiliation{Universit\'e Paris-Saclay, CNRS/IN2P3, IJCLab, 91405 Orsay, France}

\author{Yuval~Grossman}
\email{yg73@cornell.edu}
\affiliation{LEPP, Department of Physics, Cornell University, Ithaca, NY 14853, USA}

\author{Yosef Nir}
\email{yosef.nir@weizmann.ac.il}
\affiliation{Department of Particle Physics and Astrophysics, Weizmann Institute of Science, Rehovot 7610001, Israel}

\begin{abstract}
The branching fraction of the $B_s\to K^0\overline{K}{}^0$ decay has been recently measured by the LHCb and Belle experiments. We study the consistency of the measured value with three relations to other decay rates and CP asymmetries which follow from the Standard Model, and from the approximate flavor $SU(3)$ symmetry of the strong interactions. We find that each of these relations is violated at a level of above $3\sigma$. We argue that various subleading effects -- rescattering, electroweak penguins and $SU(3)$ breaking -- if larger than theoretically expected, can account for some of these puzzles, but not for all of them simultaneously.

\end{abstract}

\maketitle

\section{Introduction}
The $B_s\to K^0\overline{K}{}^0$ decay, which proceeds via the quark transition $b\to d\bar ds$, is a flavor changing neutral current (FCNC) process and, as such, constitutes a sensitive probe of new physics. There are several unique properties of this process, which makes experimental measurements of the rate and CP asymmetries highly motivated:
\begin{itemize}
\item It is a uniquely clean and rich probe of the $\bar b\to d\bar d\bar s$ transition. Another decay channel that proceeds via $\bar b\to d\bar d\bar s$ is $B^+\to\pi^+ K^0$ for which, however, CP asymmetries are not as rich. For $\bar b\to d\bar d\bar s$ two body decays where the $d\bar d$ pair bind into a meson, there is always a contribution also from the flavor changing charged current transition $\bar b\to u\bar u\bar s$.
\item If rescattering is not surprisingly large in this mode, the CP asymmetries provide clean null tests of the Standard Model (SM). Conversely, if the CP asymmetries are experimentally established, or even just bounded, we will draw important lessons about rescattering.
\item It is related by isospin to the $B_s\to K^+K^-$ decay, where the CP asymmetries have been experimentally measured.
\item It is related by U-spin, with expected only small breaking effects, to the $B^0\to   K^0\overline{K}{}^0$ decay.
\end{itemize}

From this list of features, it is clear that measuring the rate and CP asymmetries in $B_s\to K^0\overline{K}{}^0$ decay will provide new information about QCD and on new physics. Indeed, the branching fraction ${\rm BR}(B_s\to K^0\overline{K}{}^0$) was recently measured by the LHCb experiment \cite{LHCb:2020wrt} (consistent with an earlier measurement by the BELLE experiment \cite{Belle:2015gho}). In what follows, we use the currently available data on this and related $B$-meson decays and find several puzzles which further motivate an experimental effort to obtain a more precise measurement of the rate and search for the CP asymmetries in this mode.

The time-dependent CP asymmetries in the $B_s\to K^0\overline{K}{}^0$ decay have been argued to provide clean tests of the Standard Model and to probe the presence of new physics in $b\to s$ transitions in Ref.~\cite{Ciuchini:2007hx}. In Refs.~\cite{Descotes-Genon:2011rgs,Bhattacharya:2012hh}, the potential of these measurements in probing new CP violating physics in $B_s-\overline{B_s}$ mixing was analyzed. (In our work, we use the measured value of the CP violating phase as input.)  Predictions for the branching fraction and CP asymmetries in $B_s\to K^0\overline{K}{}^0$ were made using QCD factorization in Refs.~\cite{Wang:2014mua,Bobeth:2014rra}, and a global flavor-$SU(3)$ fit in Ref.~\cite{Cheng:2014rfa}.
Two recent studies of related topics in $B$-meson decays (which, however, do not incorporate $B_s\to K^0\overline{K}{}^0$ in their analysis) can be found in Refs.~\cite{Bhattacharya:2022akr,Fleischer:2022rkm}.

The plan of this paper is as follows. In Section \ref{sec:expdata} we review the experimental data that form the basis for our analysis. In Section \ref{sec:sm} we give the formalism that we use for analysing the data within the SM and list the approximations that we make. In Section \ref{sec:puzzles} we use three sets of data, each presenting a  deviation from the SM expectation at the $3\sigma$ level. In Section \ref{sec:rsewpsub} we reintroduce three effects that we neglected in the previous section, arguing that they are unlikely to provide a solution to the puzzles. In Section \ref{sec:future} we describe the improvements in the relevant measurements that can be expected in the future from the LHCb and Belle-II experiments. We conclude in Section \ref{sec:con}.

\section{Experimental data}
\label{sec:expdata}
Before we present a theoretical analysis of the $B_s\to K^0\overline{K}{}^0$ decay and the isospin and U-spin related modes, we collect in Table \ref{tab:exp} the relevant experimental information \cite{Zyla:2020zbs,Amhis:2019ckw}.

In what follows, we consider the following ratios of rates:
\beqa\label{eq:expR}
R^{ss}_{KK}&\equiv&\frac{\Gamma(B_s\to K^0\overline{K}{}^0)}{\Gamma(B_s\to K^+K^-)}=0.66\pm0.13,\no\\
R^{sd}_{KK}&\equiv&\left|\frac{V_{td}}{V_{ts}}\right|^2\frac{\Gamma(B_s\to K^0\overline{K}{}^0)}{\Gamma(B^0\to K^0\overline{K}{}^0)}=0.61\pm0.13,\no\\
R^{ud}_{\pi K}&\equiv&\frac{\Gamma(B^+\to \pi^+K^0)}{\Gamma(B^0\to \pi^-K^+)}=1.12\pm0.05,
\eeqa
where we use the measured values of the branching ratios from Table~\ref{tab:exp} and take into account the lifetimes of the various bottom mesons  \cite{Zyla:2020zbs} when translating ratios of branching ratios to ratios of rates:
\beqa
\tau(B_s)&=&(1.516\pm0.006)\times10^{-12}\ {\rm s},\no\\
\tau(B^0)&=&(1.519\pm0.004)\times10^{-12}\ {\rm s},\no\\
\tau(B^+)&=&(1.638\pm0.004)\times10^{-12}\ {\rm s}.
\eeqa

\begin{table}
\label{tab:exp}
\begin{center}
\begin{tabular}{|c|c|c||c|c| } \hline\hline
\rule{0pt}{1.2em}%
Process &  Branching ratio & Refs. & CP asymmetries & Refs. \\[2pt]\hline\hline
\rule{0pt}{1.2em}%
$B_s\to K^0\overline{K}{}^0$ & $(1.76\pm0.31)\times10^{-5}$ & \cite{Belle:2015gho,LHCb:2020wrt} & $-$ & \\
$B_s\to K^+K^-$ & $(2.66\pm0.22)\times10^{-5}$ & \cite{Belle:2010yix,CDF:2011ubb,LHCb:2012yxl} & $\begin{matrix}C^s_{K^+K^-}=+0.172\pm0.031\cr S^s_{K^+K^-}=+0.139\pm0.032\end{matrix}$& \cite{Aaij:2018tfw,LHCb:2020byh} \\
    $B^0\to K^0\overline{K}{}^0$ & $(1.21\pm0.16)\times10^{-6}$ & \cite{BaBar:2006enb,Belle:2012dmz} & $\begin{matrix}C^d_{K^0\overline{K}{}^0}=+0.0\pm0.4\cr S^d_{K^0\overline{K}{}^0}=-0.8\pm0.4\end{matrix}$& \cite{Belle:2007cga,BaBar:2006enb} \\
$B^+\to K^0\pi^+$ & $(2.37\pm0.08)\times10^{-5}$ & \cite{Belle:2012dmz,BaBar:2006enb,CLEO:2003oxc} & $A^u=-0.017\pm0.016$ & \cite{LHCb:2013vip,Belle:2012dmz,BaBar:2006enb,CLEO:2000uwz} \\
$B^0\to K^+\pi^-$ & ~$(1.96\pm0.05)\times10^{-5}$~ & \cite{Belle:2012dmz,BaBar:2006pvm,CLEO:2003oxc} & $A^d=-0.0834\pm0.0032$ ~& \cite{LHCb:2020byh,LHCb:2018pff,CDF:2014pzb,Belle:2012dmz,BaBar:2012fgk,CLEO:2000uwz} \\
\hline\hline
\end{tabular}
\caption{Experimental data from the PDG \cite{Zyla:2020zbs}. The sub-index on $C$ and $S$ represents the final state, while the super-index $u$, $d$, or $s$ corresponds to an initial $B^+$, $B^0$, or $B_s$.}
\end{center}
\end{table}

\begin{table}
\label{tab:ckm}
\begin{center}
\begin{tabular}{|c|c||c|c||c|c| } \hline\hline
\rule{0pt}{1.2em}%
Parameter &  Value & Parameter & Value & Parameter & Value \\[2pt]\hline\hline
\rule{0pt}{1.2em}%
$|V_{ub}|$ & $0.0037\pm0.0001$ & $|V_{tb}|$ & $0.99912\pm0.00004$ & $\alpha$ & $(85.2^{+4.8}_{-4.3})^o$ \\
$|V_{us}|$ & $0.2250\pm0.0007$ & $|V_{ts}|$ & $0.0411\pm0.0008$ & $\gamma$ & $(65.9^{+3.3}_{-3.5})^o$ \\
$|V_{ud}|$ & $0.9743\pm0.0002$ & $|V_{td}|$ & $0.0086\pm0.0002$ & & \\
\hline\hline
\end{tabular}
\caption{CKM parameters from the PDG \cite{Zyla:2020zbs}}
\end{center}
\end{table}

We define the CKM combinations
\beq
\lambda^{q^\prime}_{bq}=V_{q^\prime b}^*V_{q^\prime q},\ \ \ R^{bq}_{uq^\prime}=|\lambda^u_{bq}/\lambda^{q^\prime}_{bq}|\ \ \ (q^\prime=u,c,t,\ q=d,s).
\eeq
The CKM phases are defined as follows:
\beq
\gamma={\rm arg}\left(\lambda^u_{bs}/\lambda^c_{bs}\right),\qquad
\alpha={\rm arg}\left(-\lambda^t_{bd}/\lambda^u_{bd}\right).
\eeq
We will need
\beq
\lambda^u_{bs}/\lambda^t_{bs}=-R^{bs}_{ut}e^{i\gamma},\qquad
\lambda^u_{bd}/\lambda^t_{bd}=-R^{bd}_{ut}e^{-i\alpha}.
\eeq

The experimental ranges of the relevant CKM parameters are presented in Table~\ref{tab:ckm}. They lead to the following combinations which play a role in our analysis:
\beqa\label{eq:ckmexp}
&&R^{bs}_{ut}=0.0203\pm0.0007,\qquad R^{bd}_{ut}=0.420\pm0.016,\qquad |V_{td}/V_{ts}|=0.209\pm0.006,\no\\ &&
\sin\alpha\simeq1,\qquad\sin\gamma=0.91 \pm 0.02,\qquad  \cot\gamma=0.45\pm0.07.
\eeqa
We also use the following combinations:
\beqa\label{eq:scom1}
&&S^s_{K^+K^-}\cot\gamma = 0.06\pm0.02, \nonumber \\
&&2R_{bs}^{ut}\cos\gamma= 0.016\pm0.002,\nonumber \\
&&2R_{bs}^{ut}\sin\gamma= 0.036\pm0.001,\nonumber \\
&&[(C^s_{K^+K^-})^2+(S^s_{K^+K^-})^2]^{1/2}=0.22\pm0.05. 
\eeqa
%

\section{The SM: formalism and approximations}
\label{sec:sm}
In what follows we assume $SU(3)$ flavor symmetry and employ the diagrammatic approach  of Ref.~\cite{Zeppenfeld:1980ex}. 
Our starting point is the analysis of Ref.~\cite{Gronau:1994rj}. The relevant amplitudes are written as follows:
\beqa\label{eq:ampghlr}
{\cal A}(B_s\to K^0\overline{K}{}^0)&=&\lambda^t_{bs}(P+P_A),\no\\
{\cal A}(B_s\to K^+K^-)&=&-\lambda^t_{bs}(P+P_A)-\lambda^u_{bs}(T+E),\no\\
{\cal A}(B^0\to K^0\overline{K}{}^0)&=&\lambda^t_{bd}(P+P_A),\no\\
{\cal A}(B^+\to \pi^+K^0)&=&\lambda^t_{bs}P+\lambda^u_{bs}A,\no\\
{\cal A}(B^0\to \pi^-K^+)&=&-\lambda^t_{bs}P-\lambda^u_{bs}T,
\eeqa
where $P$ and $P_A$ refer to penguin and penguin annihilation diagrams, and $T$, $E$ and $A$ refer to spectator tree, exchange and annihilation diagrams.

In writing the relations in Eqs.~(\ref{eq:ampghlr})    three effects are neglected:
\begin{itemize}
\item $SU(3)$ breaking \cite{Gronau:1995hm};
\item Rescattering contributions,  $T_{RS}$ \cite{Gronau:2012gs};
\item Electroweak (EW) penguin contributions, $P_{EW}$ \cite{Gronau:1995hn}.
\end{itemize}
We discuss these effects in Section~\ref{sec:rsewpsub}.

The smallness of $R_{bs}^{ut}=|\lambda^u_{bs}/\lambda^t_{bs}|$ implies that the decays which proceed via the quark transitions $\bar b\to \bar qq\bar s$ ($q=d$ or $u$) are dominated by the gluonic penguin contributions proportional to $\lambda^t_{bs}$.  
Setting $\lambda^u_{bs} \to 0$ in Eqs.~(\ref{eq:ampghlr})
leads to the following predictions concerning CP asymmetries: 
\beq
C^s_{K^+K^-}=S^s_{K^+K^-}=0,
\eeq
and ratios of decay rates:
\beq
R^{ss}_{KK}=R^{sd}_{KK}=R^{ud}_{\pi K}=1.
\eeq
Taking into account the $\lambda^u_{bs}$ terms in Eqs.~(\ref{eq:ampghlr}) leads to small deviations from these predictions. To first order in $R^{ut}_{bs}$, we obtain, for the CP asymmetries,
\beqa\label{eq:cpvcs}
C^s_{K^+K^-}&=&2R^{ut}_{bs}\sin\gamma\times{\cal I}m[(T+E)/(P+P_A)],\\
S^s_{K^+K^-}&=&2R^{ut}_{bs}\sin\gamma\times{\cal R}e[(T+E)/(P+P_A)],
\label{eq:cpvss}
\eeqa
and for the ratios of rates,
\beqa
R^{ss}_{KK}&=&1+2R^{ut}_{bs}\cos\gamma\times{\cal R}e[(T+E)/(P+P_A)],\label{eq:rss}\\
R^{ud}_{\pi K}&=&1+2R^{ut}_{bs}\cos\gamma\times{\cal R}e[(T-A)/P],\label{eq:rud}\\
R^{sd}_{KK}&=&1.\label{eq:rsd}
\eeqa
%

\section{Puzzles involving
$B_s\to K^0\overline{K}{}^0$}
\label{sec:puzzles}
Using the above theoretical relations that assume the SM and the $SU(3)$ flavor symmetry of QCD to analyze the experimental data, we identify three puzzles involving the
$B_s\to K^0\overline{K}{}^0$ decay rate. We present them in the following subsections.

\subsection{The  $R^{sd}_{KK} =1$ puzzle.}   
\label{sec:rsdone}
$R^{sd}_{KK}$ is a ratio of two decay rates that are connected by U-spin.
One that proceed via the quark transitions $\bar b\to\bar dd\bar s$ and 
 the other one via $\bar b\to\bar ss\bar d$. 
%
They have neither tree ($T$), nor annihilation ($A$) nor exchange ($E$) contributions, hence the $R^{sd}_{KK}=1$ prediction in Eq.~(\ref{eq:rsd}). The experimental range, $R^{sd}_{KK}=0.61\pm0.13$, shows a $3\sigma$ deviation from the SM prediction. The deviation of the experimental value of $R^{sd}_{KK}$ from 1 constitutes the first puzzle.  

\subsection{The $R^{ss}_{KK}-S^s_{K^+K^-}$ puzzle} %
\label{sec:rssss}
$R^{ss}_{KK}$ is the ratio between two rates related by isospin.  Eqs.~(\ref{eq:cpvss}) and (\ref{eq:rss}) lead to the following prediction:
\beq\label{eq:rsskkskkiso}
R^{ss}_{KK}=1+S^s_{K^+K^-}\cot\gamma.
\eeq
Using Eq.~(\ref{eq:scom1}) we obtain the following range for the right hand side of Eq.~(\ref{eq:rsskkskkiso}):
\beq\label{eq:rsssskk}
1+S^s_{K^+K^-}\cot\gamma = 1.06\pm0.02.
\eeq
Thus, the experimental range, $R^{ss}_{KK} = 0.66\pm0.13$, shows a $3\sigma$ deviation from the SM prediction.  This $R^{ss}_{KK}-S^s_{K^+K^-}$ inconsistency constitutes the second puzzle.  

\subsection{The  $R^{ss}_{KK}=R^{ud}_{\pi K}$ puzzle}
\label{sec:rssrud}
Eqs.~(\ref{eq:rss}) and (\ref{eq:rud}) lead to the following relation between $R^{ss}_{KK}$ and $R^{ud}_{\pi K}$
\beq \label{eq:KK-Kpi-full}
R^{ss}_{KK}- R^{ud}_{\pi K}=2R^{ut}_{bs}\cos\gamma\times{\cal R}e[(T+E)/(P+P_A)-(T-A)/P].
\eeq
As mentioned above, it is expected that the penguin annihilation contribution is suppressed compared to the penguin contribution, $|P_A/P|\ll1$, and that the exchange and annihilation contributions are suppressed compared to the spectator tree contribution, $|E/T|\ll1$ and $|A/T|\ll1$. To first order in these small hadronic parameters, Eq.~(\ref{eq:KK-Kpi-full}) leads to the following relation:
\beq\label{eq:rssrud}
R^{ss}_{KK}/R^{ud}_{\pi K}=1+2R^{ut}_{bs}\cos\gamma\times{\cal R}e\left[(T/P)(E/T+A/T-P_A/P)\right].
\eeq

To estimate the deviation of the double ratio $R^{ss}_{KK}/R^{ud}_{\pi K}$ from unity, we first use the known values of the weak parameters to calculate $2R_{bs}^{ut}\cos\gamma\approx0.016$. The hadronic part has a large factor, $T/P$, multiplied by a small factor, $E/T+A/T-P_A/P$. As concerns $T/P$, we use Eqs.~(\ref{eq:cpvcs}) and (\ref{eq:cpvss}) to zeroth order in $|E/T|$ and $|P_A/P|$, and the values of the observables given in Eqs.~(\ref{eq:ckmexp}) and (\ref{eq:scom1}), and find
\beq\label{eq:toverp}
\left|\frac{T}{P}\right|\approx\frac{\left[(C^s_{K^+K^-})^2+(S^s_{K^+K^-})^2\right]^{1/2}}{2R^{ut}_{bs}\sin\gamma}\approx6.0\pm1.4.
\eeq
As concerns the hadronically suppressed part, while we do not assign a strict upper bound on its value, we assume that it is of order 
\beq \label{eq:small-E}
E/T+A/T-P_A/P \sim {f_B/ m_B}  \sim0.05.
\eeq
See  Ref.~\cite{Huber:2021cgk} for a recent discussion.
We thus expect
\beq
{\cal R}e\left[(T/P)(E/T+A/T-P_A/P)\right]\lsim 1.
\eeq

We conclude that the deviation of the double ratio $R^{ss}_{KK}/R^{ud}_{\pi K}$ from unity is predicted to be highly CKM-suppressed and without hadronic enhancement, and we estimate it to be of order $0.01$.
The LHS of Eq.~(\ref{eq:rssrud}) is experimentally measured to be
\beq\label{eq:rssrudexp}
R^{ss}_{KK}/R^{ud}_{\pi K}=0.59\pm0.12.
\eeq
This experimental range shows a $3.4\sigma$ deviation from the SM prediction of 1. This  $R^{ss}_{KK}-R^{ud}_{\pi K}$ inconsistency constitutes the third puzzle.

\section{Rescattering, EW penguins, and $SU(3)$ breaking}
\label{sec:rsewpsub}
As mentioned above, the analysis of Ref.~\cite{Gronau:1994rj} is conducted in the limit of $SU(3)$-flavor symmetry and neglects the contributions from rescattering and from electroweak penguins. In this section we discuss whether these missing pieces in the analysis can account for the various puzzles presented in the previous section.

\subsection{The $R^{sd}_{KK}=1$ puzzle: Rescattering}
In this subsection we argue that the $R^{sd}_{KK}=1$ puzzle cannot be explained by $SU(3)$ breaking or electroweak penguins. It can, however, be explained by rescattering.

The $B_s\to K^0\overline{K}{}^0$ and $B^0\to K^0\overline{K}{}^0$ decays are related by U-spin. Since the final state of the two decays is the same and, furthermore, does not include pions, there is no U-spin breaking proportional to a factor of $f_K/f_\pi$. The remaining effects are theoretically expected to be small, of order  $m_s/m_b$. This expectation was recently confirmed by relations between $B_s\to K^+K^-$ and $B^0\to\pi^+\pi^-$ \cite{Nir:2022bbh}. 

Given that the electromagnetic charge of the $s$ and $d$ quarks are the same, U-spin implies that the electroweak penguin contributions to  the $B_s\to K^0\overline{K}{}^0$ and $B^0\to K^0\overline{K}{}^0$ decays are also equal, and thus they do not affect the $R^{sd}_{KK}=1$ prediction.

Rescattering contributes to $B_s\to K^0\overline{K}{}^0$ via $\bar b\to\bar uu\bar s$ followed by $u\bar u\to d\bar d$. 
Rescattering contributes to $B^0\to K^0\overline{K}{}^0$ via $\bar b\to\bar uu\bar d$ followed by $u\bar u\to s\bar s$. Thus, in the presence of rescattering, Eq.~(\ref{eq:ampghlr}) is modified:
\beqa
{\cal A}(B_s\to K^0\overline{K}{}^0)&=&\lambda^t_{bs}(P+P_A)+\lambda^u_{bs}T_{RS},\no\\
{\cal A}(B^0\to K^0\overline{K}{}^0)&=&\lambda^t_{bd}(P+P_A)+\lambda^u_{bd}T_{RS}.
\eeqa
Consequently, neglecting $P_A/P$ and $R^{ut}_{bs}$ compared with $R^{ut}_{bd}$ but keeping all orders in   $T_{RS}/P$, we find
\beq
R^{sd}_{KK} = [1-2(R^{ut}_{bd}\cos\alpha-R^{ut}_{bs}\cos\gamma){\cal R}e(T_{RS}/P)+(R^{ut}_{bd}|T_{RS}/P|)^2]^{-1}.
\eeq
The experimental value of $R^{sd}_{KK}$ can be accounted for with
\beq\label{eq:trsp}
1.4\lsim|T_{RS}/P|\lsim2.4.
\eeq
Given $|T/P| \approx6$ from Eq.~(\ref{eq:toverp}), Eq.~(\ref{eq:trsp}) implies that, in order to explain the   experimental value of $R^{sd}_{KK}$, we need
\beq \label{eq:TRS-T}
|T_{RS}/T| \sim 1/3.
\eeq
While this value is somewhat large, it is not unacceptably so. It implies that rescattering is a subleading effect and can very well solve the puzzle.

If, indeed, the rescattering contribution enhances $\Gamma(B^0\to K^0\overline{K}{}^0)$ in a significant enough way to suppress $R^{sd}_{KK}$ from unity to ${\cal O}(0.6)$, then either or both (depending on the phase of $T_{RS}/P$) time-dependent CP asymmetries, $C^d_{K^0\overline{K}{}^0}$ and $S^d_{K^0\overline{K}{}^0}$, are large.
Measuring these asymmetries would thus provide a crucial test of this scenario.

For the time-dependent CP asymmetries in $B_s\to K^0\overline{K}{}^0$, we can formulate a sum rule:
\beq
\left[(C^s_{K^0\overline{K}{}^0})^2+ (S^s_{K^0\overline{K}{}^0})^2\right]^{1/2}=2R^{ut}_{bs}\sin\gamma\times|T_{RS}/P|.
\eeq
Taking into account Eq.~(\ref{eq:trsp}), we conclude that, if rescattering explains the puzzle, at least one of the CP asymmetries should be of order a few percent:
\beq
0.05 \lsim \left[(C^s_{K^0\overline{K}{}^0})^2+ (S^s_{K^0\overline{K}{}^0})^2\right]^{1/2} \lsim 0.09.
\eeq
%

\subsection{The   $R^{ss}_{KK}=1+S^s_{K^+K^-}\cot\gamma$ puzzle: Electroweak penguins}
In this subsection we argue that the $R^{ss}_{KK}=1+S^s_{K^+K^-}\cot\gamma$ puzzle cannot be explained by $SU(3)$ breaking or rescattering. It can, however, be explained by electroweak (EW) penguins, but at the cost of tension with other observables.

The $B_s\to K^0\overline{K}{}^0$ and $B_s\to K^+K^-$ decays are related by isospin. Isospin breaking is very small, of $O(0.01)$, and cannot explain the puzzle.

The value of  $|T_{RS}/T|\sim1/3$, see Eq.~(\ref{eq:TRS-T}), implies that, at best, rescattering can bring the central value of the right hand side of Eq.~(\ref{eq:rsssskk}) to 1.03, not enough to explain the puzzle.

Due to the different electromagnetic charges of the $u$ and $d$ quarks, EW penguins give different contributions to the decays in question:
\beqa
{\cal A}(B_s\to K^0\overline{K}{}^0)&=&\lambda^t_{bs}[P+P_A-(1/3)P_{EW}],\no\\
{\cal A}(B_s\to K^+K^-)&=&-\lambda^t_{bs}[P+P_A+(2/3)P_{EW}]-\lambda^u_{bs}(T+E).
\eeqa
Consequently, neglecting $|P_A/P|$ and $|E/T|$, we find 
\beqa
S^s_{K^+K^-}&=&2R^{ut}_{bs}\sin\gamma\times{\cal R}e(T/P),\no\\
R^{ss}_{KK}&=&1+2R^{ut}_{bs}\cos\gamma\times{\cal R}e(T/P)-2{\cal R}e(P_{EW}/P).
\eeqa
Thus, in the presence of EW penguins, Eq.~(\ref{eq:rsskkskkiso}) is modified:
\beq
R^{ss}_{KK}=1+S^s_{K^+K^-}\cot\gamma-2{\cal R}e(P_{EW}/P).
\eeq
To explain the puzzle we thus need
\beq
{\cal R}e(P_{EW}/P)=+0.20\pm0.07.
\eeq

While as a stand alone effect EW penguins can explain the puzzle, the required value is in contradiction with other observations.
The central value is larger by about an order of magnitude than the theoretical expectations for the color-suppressed EW penguin \cite{Gronau:1995hn}. This expectation was confirmed by an analysis of a large set of observables in $B$-meson decays to pairs of $SU(3)$-octet mesons ($\pi,K,\eta_8$)  \cite{Fleischer:2018bld}.
Furthermore, the EW penguin contributions would generate a similar shift in $R^{ud}_{\pi K}$, which is unacceptable. In fact, the experimental range, $R^{ud}_{\pi K}=1.12\pm0.05$, implies that ${\cal R}e(P_{EW}/P)=-0.03\pm0.03$.

\subsection{The $R^{ss}_{KK}=R^{ud}_{\pi K}$ puzzle: $SU(3)$ breaking}
In this subsection we argue that the  $R^{ss}_{KK}=R^{ud}_{\pi K}$ puzzle cannot be explained by rescattering or EW penguins. It is affected, in principle, by $SU(3)$ breaking, but the required size of the breaking is unacceptably large.

Neglecting $P_A/P$ and $R^{ut}_{bs}(A/P)$, we have, in the $SU(3)$ limit,
\beq\label{eq:bskkbupik}
{\cal A}(B_s\to K^0\overline{K}{}^0)={\cal A}(B^+\to\pi^+K^0).
\eeq
Neither rescattering nor EW penguins affect this equality.
Neglecting $P_A/P$ and $E/T$, we have, in the $SU(3)$ limit,
\beq\label{eq:bskkbdpik}
{\cal A}(B_s\to K^+K^-)={\cal A}(B^0\to\pi^-K^+).
\eeq
Again, neither rescattering nor EW penguins affect this equality. Hence, the $SU(3)$ prediction that $R^{ss}_{KK}\simeq R^{ud}_{\pi K}$ (up to effects that are strongly CKM suppressed), is violated by neither rescattering nor electroweak penguin contributions. 

The question is then whether $SU(3)$ breaking effects can account for the experimental result (\ref{eq:rssrudexp}). An analysis of $SU(3)$ breaking was presented in Ref.~\cite{Gronau:1995hm}. We consider the $SU(3)$ breaking effects for only the $P$ and $T$ diagrams. We neglect $P_A$, $E$ and $A$.

There are two relevant $SU(3)$-breaking diagrams related to the $P$ contributions: $P_1$ where there is a $b\to s$ transition, and $P_2$ where the $s$ quark is a spectator. Similarly, there are two relevant $SU(3)$-breaking diagrams related to the $T$ contributions: $T_1$ where there is a $W\to u\bar s$ transition, and $T_2$ where the $s$ quark is a spectator. Thus, $SU(3)$ breaking effects modify Eqs.~(\ref{eq:ampghlr}) as follows \cite{Gronau:1995hm}:
\beqa\label{eq:ampsub}
{\cal A}(B_s\to K^0\overline{K}{}^0)&=&\lambda^t_{bs}(P+P_1+P_2),\no\\
{\cal A}(B_s\to K^+K^-)&=&-\lambda^t_{bs}(P+P_1+P_2)-\lambda^u_{bs}(T+T_1+T_2),\no\\
{\cal A}(B^+\to \pi^+K^0)&=&\lambda^t_{bs}(P+P_1),\no\\
{\cal A}(B^0\to \pi^-K^+)&=&-\lambda^t_{bs}(P+P_1)-\lambda^u_{bs}(T+T_1).
\eeqa
We learn that, while each of the equalities (\ref{eq:bskkbupik}) and (\ref{eq:bskkbdpik}) is violated at order $P_2/P$, the deviation of the double ratio $R^{ss}_{KK}/R^{ud}_{\pi K}$ from unity,
\beq\label{eq:rssrudbsu}
R^{ss}_{KK}/R^{ud}_{\pi K}=1+2R^{ut}_{bs}\cos\gamma\times{\cal R}e\left\{(T/P)[(T_2/T)-(P_2/P)]\right\},
\eeq
is also CKM suppressed and thus very small.

Given that $2R_{bs}^{ut}\cos\gamma\times|T/P|\approx0.10\pm0.03$, to explain the puzzle we would need
\beq
|T_2/T-P_2/P| \gsim 3.
\eeq
We learn that, to explain the deviation of $R^{ss}_{KK}/R^{ud}_{\pi K}$ from unity,  
the $SU(3)$ breaking effect has to be unacceptably large, an order of magnitude larger than the naive expectation of $30\%$.

\section{Future prospects}
\label{sec:future}
Understanding the origin of the puzzles reported in this paper requires the analysis of additional data.  Fortunately, answers may rise from both the LHCb experiment and the Belle 2 experiment. According to Table~\ref{tab:exp}, the input measurements to this work which have the largest statistical uncertainties are ${\rm BR}(B_s\to K^0\overline{K}{}^0)$, $C^s_{K^+K^-}$ and $S^s_{K^+K^-}$. In addition, $C^s_{K^0\bar K^0}$ and $S^s_{K^0\bar K^0}$ have not been measured yet.

The Belle 2 experiment foresees to collect 50 ab$^{-1}$ at the $\Upsilon(4S)$ together with a sample at the $\Upsilon(5S)$~\cite{Belle-II:2018jsg}. 
The rapid $B_s$ oscillations make the tagging of its initial flavor impossible at Belle 2. Information regarding CP violation can, however, be derived from the study of the lifetime distribution of this decay~\cite{Belle-II:2018jsg}.

The LHCb experiment has gone through a first major upgrade \cite{LHCbUpgrade} and a second one is foreseen for 2030~\cite{LHCbUpgrade2}. The integrated luminosity that is expected to be reached is 23 fb$^{-1}$ (Run 1-3) for the first and 300 fb$^{-1}$ for the second upgrade (Run 1-5).

At LHCb, the $B_s$ oscillation can be resolved, as demonstrated in Ref~\cite{LHCb:2021moh}. While the branching fraction sensitivity can be directly estimated from the expected yields, the sensitivity on CP asymmetries must be extrapolated from the analogous decays $B_s\to K^+K^-$ and $B^0\to \pi^+\pi^-$~\cite{LHCb:2018ro}.

\subsection{The rates: BR($B^0\to K^0\overline{K}{}^0$) and BR($B_s\to K^0\overline{K}{}^0$)}
Starting from the yields quoted in Ref.~\cite {LHCb:2020wrt} and assuming the same scaling adopted in Ref.~\cite{LHCb:2018ro}, a simple back of the envelope estimate gives yields of about 1500 $B^0$, and 4800 $B_s$ decaying to a $K_S K_S$ final state with 300 fb$^{-1}$. Assuming that background dilution is negligible, one can expect to reach a branching fraction precision of 0.09 (0.026) for the $B^0$ decay with 23 (300) fb$^{-1}$, and of 0.05 (0.014) for the  $B_s^0$ decay with 23 (300) fb$^{-1}$. The improvement of the precision on BR($B_s\to K^0\overline{K}{}^0$) from the current 0.18 by a factor of 3.5 (12) will be of impact for all three puzzles.

\subsection{The CP asymmetries for the neutral modes: $C^{d,s}_{K_SK_S}$ and $S^{d,s}_{K_SK_S}$}

The fairly sizable samples also open the possibility to access the CP observables $C_{K_SK_S}$ and $S_{K_SK_S}$ in both the $B^0$ and  $B_s^0$ systems. In order to estimate this, we assume that the achievable precision on the $C_{X}$ and $S_{X}$ parameters is equal for all $B^0_{(s)}$ decays given equal signal yields. Furthermore, we assume that the flavor tagging performances are roughly the same for all hadronic $B^0_{(s)}$ decays at LHCb. Then one can scale the sensitivities reported for $B_s\to K^+K^-$ and $B^0\to \pi^+\pi^-$ in ~\cite{LHCb:2018ro} by the expected yields at a given luminosity. This computation leads to an expected precision of 0.89 and 0.48 for the $B^0$ and  $B_s^0$ decays, respectively, with 23 fb$^{-1}$. The precision on these quantities is expected to improve to 0.25 and 0.13 with 300 fb$^{-1}$. A better estimate of these extrapolations and assumptions will be possible once LHCb will explore the Run 3 data.  
Though challenging to measure, this information is key to address the $R^{sd}_{KK}=1$ puzzle.

\subsection{The CP asymmetries for the charged  modes: $C^s_{K^+K^-}$ and $S^s_{K^+K^-}$}

Ref.~\cite{LHCb:2018ro} provides an extrapolation of the statistical sensitivity for the CP violating parameters of the $B_s\to K^+K^-$ decay at LHCb.   It is worth mentioning that the scaling used for these extrapolations is conservative, given that the performances of  the flavor tagging, the decay-time resolution and the particle identification performance were assumed to be the same as in Run 1. One can expect to reach already precision of 0.015 for each of $C^s_{K^+K^-}$ and $S^s_{K^+K^-}$ with Run 1-3 data.The expected precision improves to  0.004 for the same observables with 300 fb$^{-1}$.
This will contribute to shedding the light on the second and third puzzles. 

\section{Discussion and Conclusions}
\label{sec:con}
Our starting point is the measurement of ${\rm BR}(B_s\to K^0\overline{K}{}^0)$ by the LHCb \cite{LHCb:2020wrt} and Belle \cite{Belle:2015gho} experiments. Our analysis involves branching fractions and CP asymmetries in four additional $B$-meson decays related to $B_s\to K^0\overline{K}{}^0$ by $SU(3)$-flavor symmetry: $B_s\to K^+K^-$, $B^0\to K^0\overline{K}{}^0$, $B^+\to K^0\pi^+$ and $B^0\to K^+\pi^-$.  
Our analysis demonstrates that the values of the CP asymmetries $S^s_{K^+K^-}$ and $C^s_{K^+K^-}$, and of the ratios of rates 
\beq
R^{sd}_{KK}=\left|\frac{V_{td}}{V_{ts}}\right|^2\frac{\Gamma(B_s\to K^0\overline{K}{}^0)}{\Gamma(B^0\to K^0\overline{K}{}^0)}, \qquad R^{ud}_{\pi K}=\frac{\Gamma(B^+\to \pi^+K^0)}{\Gamma(B^0\to \pi^-K^+)},
\eeq 
can be accounted for with the following hierarchy of contributions to $b\to s$ transitions:
\begin{itemize}
\item A dominant contribution from gluonic penguin, proportional to $V_{tb}^*V_{ts}$.
\item Tree level contribution of ${\cal O}(0.06)$ of the leading gluonic penguin to $B_s\to K^+K^-$.
\item Rescattering contribution of ${\cal O}(0.03)$ of the leading gluonic penguin contribution to $B_s\to K^0\overline{K}{}^0$.
\item Color suppressed electroweak penguin contributions of ${\cal O}(0.03)$ of the leading gluonic penguin contributions to  $B_s\to K^0\overline{K}{}^0$ and to $B_s\to K^+K^-$.
\end{itemize}
These contributions cannot, however, explain the low value of 
\beq
R^{ss}_{KK}=\frac{\Gamma(B_s\to K^0\overline{K}{}^0)}{\Gamma(B_s\to K^+K^-)}
\eeq
compared to unity and, even more so, compared to $R^{ud}_{\pi K}$. In fact, they imply that $R^{ss}_{KK}\gsim1$ . The discrepancy is at the $3\sigma$ level. 

The large deviation of $R^{ss}_{KK}$ from unity and/or from $R^{ud}_{\pi k}$ is the combined puzzle. It cannot be accounted for even after considering various effects -- rescattering, color-suppressed EW penguins and $SU(3)$ breaking -- that are expected to be small. 

While we did not look for possible explanations of the puzzles, 
we note that if ${\rm BR}(B_s\to K^0\overline{K}{}^0)$ would turn out to be $3\sigma$ higher than its experimental central value, all three puzzles that we presented will be solved, and the situation would be consistent with the expectation that rescattering and color-suppressed electroweak penguins give very small contributions to the decays in question.

The puzzles described in this work call for an experimental effort to improve the accuracy of the relevant measurements. In particular, searching for CP asymmetries in $B_s\to K^0\overline{K}{}^0$ and in $B^0\to K^0\overline{K}{}^0$ might shed light on the solution(s) to these puzzles.

\section*{Acknowledgements} 
We are grateful to  L.~Henry, M.~Kenzie, S.~Perazzini, and S. Schacht for fruitful discussions. 
The work of YG and YN is supported by the United States-Israel Binational Science Foundation (BSF), Jerusalem, Israel (grant number 2018257).
The work of YG is supported in part by the NSF grant PHY1316222. 
YN is the Amos de-Shalit chair of theoretical physics, and is supported by grants from the Israel Science Foundation (grant number 1124/20),  the Minerva Foundation (with funding from the Federal Ministry for Education and Research), and by the Yeda-Sela (YeS) Center for Basic Research.


\end{document}